\documentstyle [12pt] {article}

\catcode`@=12
\begin{document}
\thispagestyle{empty}
\begin{flushright} UCRHEP-T291\\November 2000\
\end{flushright}
\vspace{0.5in}
\begin{center}
{\Large \bf Naturally Small Seesaw Neutrino Mass with\\
No New Physics Beyond the TeV Scale\\}
\vspace{1.5in}
{\bf Ernest Ma\\}
{\sl Department of Physics\\}
{\sl University of California\\}
{\sl Riverside, CA 92521\\}
\vspace{1.5in}
\end{center}
\begin{abstract}\
If there is no new physics beyond the TeV energy scale, such as in a theory 
of large extra dimensions, the smallness of the seesaw neutrino mass, i.e. 
$m_\nu = m_D^2/m_N$, cannot be explained by a very large $m_N$.  In contrast 
to previous attempts to find an alternative mechanism for a small $m_\nu$, 
I show how a solution may be obtained in a simple extension of the Standard 
Model, without using any ingredient supplied by the large extra dimensions.
It is also experimentally testable at future accelerators.
\end{abstract}

\newpage
\baselineskip 24pt

In the minimal Standard Model of particle interactions, neutrinos are 
massless but they may acquire naturally small Majorana masses through the 
effective dimension-five operator \cite{wema}
\begin{equation}
{1 \over \Lambda} (\nu_i \phi^0 - l_i \phi^+)(\nu_j \phi^0 - l_j \phi^+),
\end{equation}
where $\Lambda$ is an effective large mass scale, and $\Phi = (\phi^+,\phi^0)$ 
is the usual Higgs doublet with a nonzero vacuum expectation value, 
$\langle \phi^0 \rangle = v$.  The most common realization of this operator 
is the canonical seesaw mechanism \cite{seesaw}, where three heavy 
(right-handed) singlet neutrinos $N_i$ are introduced so that
\begin{equation}
m_\nu = {m_D^2 \over m_N},
\end{equation}
with $m_D = fv$, hence $\Lambda = m_N/f^2$ in Eq.~(1).  Given that $m_\nu$ 
is at most of order 1 eV and $f$ should not be too small, the usual thinking 
is that $m_N$ has to be very large, i.e. $m_N >> v$.  As such, this famous 
mechanism must be accepted on faith, because there cannot be any direct 
experimental test of its validity.

Consider now the possibility that there is no new physics beyond the TeV 
energy scale.  This is an intriguing idea proposed recently in theories of 
large extra dimensions \cite{exdim}.  Since a large $m_N$ is not available, 
the smallness of $m_\nu$ in such theories is usually accomplished 
\cite{exnu,exnus} by putting $N$ in the bulk and then pairing it with $\nu$ 
to form a \underline {Dirac} neutrino so that its mass is suppressed by the 
volume of the extra dimensions.  Another approach is to break lepton number 
spontaneously in the bulk through a scalar singlet \cite{shine} which 
``shines'' in our world as a small vacuum expectation value.  This mechanism 
may then be combined with the triplet Higgs model of \underline {Majorana} 
neutrino mass \cite{masa} to allow direct experimental determination of the 
relative magnitude of each element of the neutrino mass matrix from $\xi^{++} 
\to l^+_i l^+_j$ decay \cite{marasa}.

Instead of using an ingredient supplied by the large extra dimensions, I 
show in the following how Eq.~(2) may be realized naturally with $m_N$ of 
order 1 TeV in a simple extension of the Standard Model.  This means that 
$m_D$ should be small, i.e. $m_D << 10^2$ GeV.  If it comes from $\phi^0$ 
as in the Standard Model, that would not be natural; but as shown below, it 
will come instead from another doublet with a naturally small vacuum 
expectation value \cite{lima}.  This new realization of the seesaw mechanism 
will allow direct experimental tests of its validity, as discussed below.

Consider the minimal Standard Model with three lepton families:
\begin{equation}
\left( \begin{array}{c} \nu_i \\ l_i \end{array} \right)_L \sim (1,2,-1/2), 
~~~ l_{iR} \sim (1,1,-1),
\end{equation}
where their transformations under the standard $SU(3)_C \times SU(2)_L \times 
U(1)_Y$ gauge group are denoted as well.  I now add three neutral fermion 
singlets
\begin{equation}
N_{iR} \sim (1,1,0),
\end{equation}
but instead of assigning them lepton number $L=1$, so they can pair up 
with the lepton doublet through the interaction $\bar N_R (\nu_L \phi^0 - 
l_L \phi^+)$, I assign them $L=0$ to forbid this Yukawa term.  To complete 
my model, a new scalar doublet
\begin{equation}
\left( \begin{array}{c} \eta^+ \\ \eta^0 \end{array} \right) \sim (1,2,1/2)
\end{equation}
is introduced with lepton number $L=-1$.  Hence the terms
\begin{equation}
{1 \over 2} M_i N_{iR}^2 + f_{ij} \bar N_{iR} (\nu_{jL} \eta^0 - l_{jL} 
\eta^+) + h.c.
\end{equation}
appear in the Lagrangian.  The effective operator of Eq.~(1) for neutrino 
mass is then replaced by one with $\eta$ instead of $\phi$, and if 
$\langle \eta^0 \rangle = u$ is naturally small, the corresponding scale 
$\Lambda$ will not have to be so large and $M_i$ of Eq.~(6) may indeed be 
of order 1 TeV.

The Higgs potential of this model is given by
\begin{eqnarray}
V &=& m_1^2 \Phi^\dagger \Phi + m_2^2 \eta^\dagger \eta + {1 \over 2} 
\lambda_1  (\Phi^\dagger \Phi)^2 + {1 \over 2} \lambda_2 (\eta^\dagger \eta)^2 
\nonumber \\ && + \lambda_3 (\Phi^\dagger \Phi)(\eta^\dagger \eta) + \lambda_4 
(\Phi^\dagger \eta)(\eta^\dagger \Phi) + \mu_{12}^2 (\Phi^\dagger \eta 
+ \eta^\dagger \Phi),
\end{eqnarray}
where the $\mu_{12}^2$ term breaks $L$ explicitly but softly \cite{ma91}. 
Note that given the particle content of this model, the $\mu_{12}^2$ term 
is the only possible soft term which also breaks $L$.

For $\langle \phi^0 \rangle = v$ and $\langle \eta^0 \rangle = u$, the 
equations of constraint are
\begin{eqnarray}
v[m_1^2 + \lambda_1 v^2 + (\lambda_3 + \lambda_4) u^2] + \mu_{12}^2 u &=& 0, 
\\ u[m_2^2 + \lambda_2 u^2 + (\lambda_3 + \lambda_4) v^2] + \mu_{12}^2 v 
&=& 0.
\end{eqnarray}
Consider the case
\begin{equation}
m_1^2 < 0, ~~~ m_2^2 > 0, ~~~ |\mu_{12}^2| << m_2^2,
\end{equation}
then
\begin{equation}
v^2 \simeq {-m_1^2 \over \lambda_1}, ~~~ u \simeq {-\mu_{12}^2 v \over 
m_2^2 + (\lambda_3 + \lambda_4) v^2}.
\end{equation}
Hence $u$ may be very small compared to $v (=174$ GeV).  For example, if $m_2 
\sim 1$ TeV, $|\mu_{12}^2| \sim 10$ GeV$^2$, then $u \sim 1$ MeV.  The 
relative smallness of $|\mu_{12}^2|$ may be attributed to the fact that it 
corresponds to the explicit breaking of lepton number in $V$ of Eq.~(7). 
[The usual argument here is that if $|\mu_{12}^2|$ were zero, then the 
model's symmetry is increased, i.e. lepton number would be unbroken. 
Hence the assumption that it is small compared to $|m_1^2|$ or $m_2^2$ is 
``natural''.  If $|\mu_{12}^2|$ were much larger, then $u$ would be 
proportionally larger, and since $m_\nu$ scales as $u^2$, neutrino masses 
would be too large.  It would also mean that the two scalar doublets mix 
to a substantial degree, which is not the case here, as discussed later 
in the paper.  If $|\mu_{12}^2|$ were much smaller, then neutrino masses 
would be too small to account for the present observation of neutrino 
oscillations.]

The $6 \times 6$ mass matrix spanning $[\nu_e, \nu_\mu, \nu_\tau, N_1, N_2, 
N_3]$ is now given by
\begin{equation}
{\cal M}_\nu = \left[ \begin{array} 
{c@{\quad}c@{\quad}c@{\quad}c@{\quad}c@{\quad}c} 0 & 0 & 0 & f_{e1}u & 
f_{e2}u & f_{e3}u \\ 0 & 0 & 0 & f_{\mu1}u & f_{\mu_2}u & f_{\mu3}u \\ 0 & 0 
& 0 & f_{\tau1}u & f_{\tau2}u & f_{\tau3}u \\ f_{e1}u & f_{\mu1}u & 
f_{\tau1}u & M_1 & 0 & 0 \\ f_{e2}u & f_{\mu2}u & f_{\tau2}u & 0 & M_2 & 0 \\ 
f_{e3}u & f_{\mu3}u & f_{\tau3}u & 0 & 0 & M_3 \end{array} \right].
\end{equation}
The mixing between $\nu$ and $N$ is thus of order $fu/M$, which will allow 
the physical $N$ to decay through its small component of $\nu$ to 
$l^\pm W^\mp$.  The effective mass matrix spanning the three light neutrinos 
is then
\begin{equation}
{\cal M}_{ij} = \sum_k {f_{ik} f_{jk} u^2 \over M_k}
\end{equation}
and is of order 1 eV if $f$ is of order unity. 

There are five physical Higgs bosons:
\begin{eqnarray}
&& h^\pm = {v \eta^\pm - u \phi^\pm \over \sqrt {v^2 + u^2}}, ~~~ A = 
{\sqrt{2} (v {\rm Im} \eta^0 - u {\rm Im} \phi^0) \over \sqrt {v^2 + u^2}}, 
\\ && h_1^0 \simeq {\sqrt{2} (v {\rm Re} \phi^0 + u {\rm Re} \eta^0) \over 
\sqrt {v^2 + u^2}}, ~~~ h_2^0 \simeq {\sqrt{2} (v {\rm Re} \eta^0 - u {\rm Re} 
\phi^0) \over \sqrt {v^2 + u^2}},
\end{eqnarray}
with masses given by
\begin{eqnarray}
m^2_{h^\pm} &=& m_2^2 + \lambda_3 v^2 + (\lambda_2 - \lambda_4) u^2 - 
\mu_{12}^2 u/v, \\ m^2_A &=& m_2^2 + (\lambda_3 + \lambda_4) v^2 + \lambda_2 
u^2 - \mu_{12}^2 u/v, \\ m^2_{h_1^0} &=& 2 \lambda_1 v^2 + {\cal O}(u^2), \\ 
m^2_{h_2^0} &=& m_2^2 + (\lambda_3 + \lambda_4) v^2 + {\cal O}(u^2).
\end{eqnarray}
From Eq.~(15), it is clear that $h_1^0$ behaves very much like the 
standard Higgs boson, as far as its coupling to all other particles are 
concerned.  The new scalar particles of this model, i.e. $h^\pm$, $A$, and 
$h_2^0$ (all with mass $\sim m_2$), as well as $N_{iR}$ are now also 
accessible to direct experimental discovery in future accelerators.  The key 
is of course Eq.~(6).

Consider first the case $m_2 > M_i$.  The decay chain
\begin{equation}
h^+ \to l^+_i N_j, ~~~ {\rm then} ~~ N_j \to l_k^\pm W^\mp,
\end{equation}
will determine the relative magnitude of each element of ${\cal M}_\nu$ 
in Eq.~(12).  Note that $h^+ \to l_i^+ l_k^+ W^-$ can be a very distinct 
experimental signature.  This direct test of the seesaw mechanism as 
the source of neutrino mass will remove all uncertainties regarding the 
indirect determination of ${\cal M}_\nu$ from neutrino-oscillation 
experiments.

Whereas $h^\pm$ is readily produced through its electromagnetic interaction, 
$h_2^0$ and $A$ are only produced through their weak interactions, i.e. 
$Z \to h_2^0 A$ and $W^\pm \to h^\pm (h_2^0,A)$.  Their decay chain
\begin{equation}
h_2^0, A \to \nu N, ~~~ {\rm then} ~~ N \to l^\pm W^\mp,
\end{equation}
is also less informative because the flavor of the neutrino in the first decay 
product cannot be identified experimentally.

Consider now the case $M_i > m_2$. The decay
\begin{equation}
N_i \to l_j^\pm h^\mp
\end{equation}
will determine $|f_{ij}|$ in Eqs.~(6) and (12).  The subsequent decay of 
$h^\pm$ occurs through its small component of $\phi^\pm$, so it is dominated 
by the final states $t \bar b$ or $\bar t b$ and should be easily 
identifiable.  The production of $N$ in a hadron collider is difficult, but 
with an $e^+ e^-$ or $\mu^+ \mu^-$ collider, it can be produced easily in 
pairs through $h^\pm$ exchange.  The decay of the two $N$'s will include 
final states of the type $l_i^+ l_j^+ b b \bar t \bar t$ which are very 
distinctive.  Note that whether $m_2 > M_i$ or $M_i > m_2$, $e^+e^-$ or 
$\mu^+\mu^-$ production of $N$ is possible.  In the former case, $N$ decays 
into $l^\pm W^\mp$, $\nu Z$, and $\bar \nu Z$, whereas in the latter case, 
it decays into $l^\pm h^\mp$, $\nu h_2^0$, $\bar \nu h_2^0$, $\nu A$, and 
$\bar \nu A$ (with $h_2^0$ and $A$ both decaying into $\bar t t$).  Either 
possibility will allow the experimenter to determine $|f_{ij}|$ and $M_i$, 
thereby obtaining the neutrino mass matrix up to an overall scale factor.

Lepton flavor violation (LFV) is a generic feature of all models of neutrino 
mass.  In this model, there is no LFV at tree level for charged leptons.  
However, it does occur in one loop through $\eta$ and $N$ exchange.  
The extra scalar doublet $(\eta^+, \eta^0)$ also contributes to the oblique 
parameters in precision electroweak measurements \cite{hahama}.  These 
contributions are easily calculated \cite{kema}. For example, with 
$m_2^2 >> M_Z^2$,
\begin{eqnarray}
\Delta S &=& {1 \over 24 \pi} {\lambda_4 v^2 \over m_2^2}, \\ 
\Delta T &=& {1 \over 96 \pi} {1 \over s^2 c^2 M_Z^2} {\lambda_4^2 v^2 \over 
m_2^2},
\end{eqnarray}
where $s^2 = \sin^2 \theta_W$, $c^2 = \cos^2 \theta_W$. They are clearly 
negligible and will not change the excellent experimental fit of the 
minimal Standard Model.

In summary, a new seesaw model of neutrino mass is proposed, where a second 
scalar doublet $(\eta^+,\eta^0)$ with lepton number $L=-1$ is added to the 
minimal Standard Model together with three neutral right-handed fermion 
singlets $N_i$ with lepton number $L=0$.  Thus $N_i$ is allowed to have a 
Majorana mass $M_i$ as well as the interaction $f_{ij} \bar N_{iR} (\nu_{jL} 
\eta^0 - l_{jL} \eta^+)$.  Hence $m_\nu$ is proportional to $\langle \eta^0 
\rangle^2 /M_i$ and if $\langle \eta^0 \rangle << \langle \phi^0 \rangle$, 
$M_i$ may be of order 1 TeV and be observable experimentally.  This is 
accomplished with the Higgs potential of Eq.~(7) where $L$ is broken 
explicitly and uniquely with the soft term $\Phi^\dagger \eta + \eta^\dagger 
\Phi$.

The decay of $N_i$ into a charged lepton together with a charged Higgs boson 
or $W$ boson will determine the relative magnitude of each element of the 
neutrino mass matrix.  Just as the discovery of the standard Higgs boson 
would settle the question of how quarks and leptons acquire mass, the 
discovery of $N_i$ and the new scalar doublet of this model would settle the 
question of how neutrinos acquire mass, and remove all uncertainties regarding 
the indirect determination of ${\cal M}_\nu$ from neutrino-oscillation 
experiments.

This work was supported in part by the U.~S.~Department of Energy under 
Grant No.~DE-FG03-94ER40837.

\bibliographystyle{unsrt}

\end{document}